\documentstyle[12pt,aasms4]{article}
\singlespace
\epsfverbosetrue
\def\teff{{\it T$_{\rm eff}$}}
\def\logg{{\rm log~g}}

\def\kms{{km~s$^{-1}$}}  
\def\etal{{\rm et~al.\,}}

\begin{document}

\title{ A-type Supergiant Abundances in the SMC: 
 Probes of Evolution }
\author{ Kim A.\, Venn\altaffilmark{1}}
\affil{Macalester College} 
\authoraddr{1600 Grand Ave., St.\,Paul, MN, 55105}

\altaffiltext{1}{Postdoctoral work while at
  Universit\"at Sternwarte M\"unchen, 
    Scheinerstrasse 1, D-81679, Munich, Germany, and  
  Max-Planck-Institut f\"ur Astrophysik
    Karl-Schwarzschildstr.~2, 85740 Garching, Germany}

\begin{abstract}

New abundances of N, O, Na, Mg, Si, Ca, Sc, Ti, Cr, Fe, Sr, Zr, and Ba
are presented for 10 A-type supergiants in the SMC, plus upper limits
for C.  In interpreting the CNO results for constraints on stellar evolution
theories, careful attention has been paid to the comparison abundances, 
i.e., the present day abundances of SMC nebulae and B-dwarf stars.
These new results are also compared to published results from F-K supergiant 
analyses, and found to be in good agreement when both sets of data are carefully 
examined as differential (SMC minus Galactic standard) abundances. 

With the exception of nitrogen, very small star-to-star abundance 
variations are found for all elements in this analysis. 
The N variations are {\it not} predicted by standard stellar 
evolution models.    Instead, the results support the new 
predictions reported from rotating stellar models,  where the range
in nitrogen is the result of partial mixing of CN-cycled gas 
from the stellar interior due to main-sequence rotation at different
rates (c.f., Langer \& Heger 1998).   The overall overabundance 
of nitrogen in the sampled stars also implies these stars have undergone 
the first dredge-up in addition to having been mixed while on the 
main-sequence.

The alpha-elements (O, Mg, Si, Ca, Ti) have similar underabundances
to Fe, which is not the same as seen in metal-poor stars in the solar 
neighborhood of the Galaxy.    In addition, certain light 
s-process elements (Zr, Ba) are slightly  more underabundant than Fe, 
which is predicted by the bursting chemical evolution model presented 
by Pagel \& Tautvai\u{s}ien\.e (1998) for the SMC.

\end{abstract}

\keywords{ stars: abundances --  stars: atmospheres --
stars: evolution -- supergiants -- Small Magellanic Cloud} 

\section{Introduction} 

Abundance analyses of supergiants in the Small Magellanic Cloud (SMC)
have several applications.   In particular, they can be used to 
study metallicity effects on the evolution of massive stars, and
they yield information on the present-day chemical composition of the 
SMC, which is an important constraint on SMC evolution models. 

The A-type supergiants have proved to be very useful abundance
indicators based on the tailored analyses of 22 Galactic stars
by Venn (1995a,b).    In fact, the accuracy of the elemental abundances
from a tailored analysis (see Section 3 for details) certainly rivals
that from any other supergiant analysis, e.g., F-K supergiants.
In A-type supergiant atmospheres, it is now known that NLTE effects
are often {\it less} significant on the model atmosphere structure 
(temperature and pressure versus optical depth) than line blanketing 
effects (Przybilla 1997, also see Section 3.1 for a demonstration 
using H$\gamma$ line profiles).   In addition, the spectra of A-type 
supergiants are very clean (unlike the more crowded spectra of F-K
supergiants), with many unblended, important spectral 
features that are ideal abundance indicators.
 
A-type supergiants are also known to be the brightest stars at visual 
wavelengths, which makes them ideal targets for extragalactic stellar 
analyses.   It is possible to observe features from a wide variety of 
elements in their spectra, ranging from light elements (B, CNO) 
to s-process elements (Sr, Zr, Ba).
These can be used to study abundance gradients or element abundance
dispersions in Local Group galaxies (e.g., McCarthy \etal 1995,
Venn \etal 1998).    
In the SMC, elemental abundances from F-K supergiants 
(Luck \etal 1998, Hill \etal 1997a,b, Luck \& Lambert 1992, 
Spite \etal 1989a,b, Russell \& Bessell 1989), 
planetary nebulae (c.f., Dopita \etal 1997), 
supernovae remnants (c.f., Russell \& Dopita 1990), 
and H~II regions (Dufour 1984, Russell \& Dopita 1990, 
Garnett \etal 1995, Kurt \etal 1998)
have shown that $\alpha$-elements share the underabundance
of iron-group elements.   This is interesting because metal-poor stars
in the Galaxy tend to have higher abundances of $\alpha$ elements 
(c.f., Edvardsson \etal 1993) 
due to early enrichments from Type II supernovae
(c.f., Pagel \& Tautvai\u{s}ien\.e 1995).
Various SMC evolution scenarios have been proposed to account 
for these observations
(e.g., Pagel \& Tautvai\u{s}ien\.e 1998, Tsujimoto \etal 1995, 
Russell \& Dopita 1992).   The abundances available from the 
A-type supergiants nicely complement those in the literature
for these studies. 

That A-supergiants are evolved massive stars also means that the 
light element abundances can be used to study stellar evolution effects. 
In the Galaxy, most supergiants show evidence for some mixing of
CNO-cycled gas at their surfaces (e.g., Gies \& Lambert 1992, 
Lennon 1994, Venn 1995a, Luck \& Lambert 1985, Spite \& Spite 1990, 
Boyarchuk \etal 1985), an observation that most stellar 
evolution scenarios fail to predict.   Additionally, some main-sequence
B stars show enhanced N (Gies \& Lambert 1992), which strongly suggests
that slight mixing can occur on the main-sequence in a small number of 
massive stars.  New massive star evolution calculations that include rotation
effects do predict some mixing between the stellar layers which can 
cause slight changes in the surface abundances 
(e.g., Langer \& Heger 1998, Talon \etal 1998, Maeder \& Zahn 1998, Meynet 1997,
Fliegner \etal 1996, Denissenkov 1994). 
The extent of mixing depends on several parameters, 
e.g., rotation rate and mass, but the success of mixing can 
also depend on convection parameters, mass loss rate, and metallicity.
The effects of rotation on stellar evolution calculations may 
have far reaching consequences, e.g., the evolution track for 
a rapidly-rotating 10 M$_\odot$ star can intersect that of a 
slowly-rotating 20 M$_\odot$ star in the blue supergiant region
(see Fig.~2 in Fliegner \etal 1996).   Such a change in evolution 
tracks could have a great impact on all uses of stellar evolution models, 
including mass and age estimates of young stars and clusters and
population syntheses of star-forming regions and galaxies.
The Magellanic Clouds provide an ideal laboratory to study the 
metallicity effects of these new stellar evolution models.

In this paper, new atmospheric parameters and elemental abundances 
are determined for 10 SMC A-type supergiants.  The results are used 
to examine metallicity effects on stellar evolution through comparisons 
with Galactic A-type supergiants (to reduce systematic errors) 
and comparisons with SMC nebular and B-dwarf abundances.    
In addition, the abundances provide new data to constrain 
chemical evolution models of the SMC, the closest and best 
studied of all metal-poor dwarf irregular galaxies. 

\section{Observations}

Spectral data were taken during two observing runs at the European
Southern Observatory in September and November 1995 at the 3.6~m
telescope.   CASPEC (R$\sim$30,000) was used in two wavelength regions, 
providing full wavelength coverage from 3900 to 8300~\AA, with only 
a small gap from 5300 to 5900~\AA.   The S/N ranged from 70 to 120 
per resolution element, and rapidly-rotating hot stars with similar 
air masses to the SMC were observed at red wavelengths to remove 
telluric lines.    Standard 
IRAF\footnote{IRAF is distributed by the National
Optical Astronomy Observatories which is operated by the 
Association of Universities for Research in Astronomy, Inc.,
under cooperative agreement with the National Science Foundation.}
packages were used for the 
data reduction (for more details see Venn 1993). 

Program stars were choosen from the Azzopardi \& Vigneau (1982)
catalogue of SMC bright stars.   This catalogue provided positions,
visual magnitudes, colors, and spectral types with luminosity classes.
An attempt has been made to observe A-supergiants over a range of
luminosities, thus masses, especially for comparison with the
Galactic stars; see Fig.~1.  
Furthermore, lower luminosity A-type supergiants 
(e.g., with luminosity classes Ib) are not expected to suffer significantly 
from microvariability, stellar wind effects, or departures from LTE (for
lines that form deep in the photosphere).   Therefore, analysis of those
stars should yield the most reliable abundance results, although in 
the SMC (see below) as in the Galaxy (Venn 1995 a,b), no significant
differences in abundances have been seen between the Ia and Ib supergiants.

\section{ Atmospheric Analysis}

Kurucz ATLAS9 (1979, 1991) model atmospheres have been adopted throughout.
The assumptions of LTE and plane-parallel geometry for the model structure 
have been shown to be satisfactory for A-type supergiants in a {\it tailored} 
analysis, i.e., when only weak lines that form deep in the atmosphere are 
used (see the discussion by Venn 1995b).
The important advantage of the ATLAS9 models is the extensive metal line 
blanketing which has a significant effect on the temperature structure
of A-type supergiants. 

As an example, an investigation of the effects of NLTE on the
H$\gamma$ line in A-supergiant model atmospheres shows that 
the results from the ATLAS9 LTE models are very similar to those 
from the Kunze NLTE models (Kudritzki, private communications) at 
low gravities.  In Fig.2, H$\gamma$ line profiles are shown based 
on ATLAS9 models (with extensive line blanketing via an opacity 
distribution function, ODF), ATLAS8 models (no line blanketing, H and He 
bound-free edges only), and Kunze NLTE models (no line blanketing, 
H and He bound-free edges only). 
At the lowest gravities, all three profiles are very similar, 
e.g., at \teff=8800~K and \logg=1.1, (see Fig.~2, left panel).    
Thus, NLTE and line blanketing are equally unimportant!
At higher gravities, the profiles do not agree as well, yet we 
expect the Kurucz models to be better since NLTE effects 
{\it decrease} with increasing gravity and line blanketing effects 
remain important.  Thus, Kurucz ATLAS9 line-blanketed model atmospheres 
are the best available for A-supergiant abundance analyses at present.

In addition, Kaufer \etal (1996) have shown that some very luminous 
A-supergiants have strongly variable Balmer lines, which would be a
problem for this analysis.    However, this variability is predicted 
to be due to changes in a stellar wind, and does not appear to affect 
the photospheric absorption lines, including the extended H$\gamma$ 
line wings.

\subsection {Atmospheric Parameters}

The atmospheric parameters determined for the SMC A-type supergiants 
are listed in Table~1.   The ``new'' spectral types are based on the
atmospheric parameters as listed in Table~1 (``old'' spectral types
are those listed by Azzopardi \& Vigneau 1982). 
These stars were analysed using the same 
methods as Venn (1995a,b).   

Atmospheric parameters were determined 
from the intersection point of two spectroscopic parameters that are 
sensitive to both temperature and gravity.   One indicator is the 
wing profile of the H$\gamma$ (and often H$\delta$) Balmer line.  
Sample profile fits are shown in Fig.~3.
The wings of the Balmer lines form deep in the atmosphere where
uncertainties due to recombination in a stellar wind or extended envelope
and microvariability effects can be ignored.   
The other is the locus of parameters that yield 
log(Mg~I/H)=log(Mg~II/H) (and where the Mg abundance itself 
is not fixed).   Sample spectra 
of some Mg~I and Mg~II absorption lines are shown in Fig.~4. 
The accuracies in the atmospheric parameters resulting from these
spectral indictors are listed in Table~2. 

If only weak lines of Mg~I and Mg~II are used, that form deep in the 
photosphere, then NLTE effects on the line formation are small.
Calculations show that NLTE corrections are typically less than 
0.1~dex throughout when the line strengths are less than about 
100 m\AA (using the detailed Mg~I/II model atom by Gigas 1988, and
analysis methods by Venn 1995b).
NLTE corrections have been included here when determining the atmospheric
parameters, even though the corrections are small (typically +0.1~dex for 
Mg~I and $-$0.1~dex for Mg~II for weak lines only).  Estimates
of the systematic offsets if NLTE is not considered are shown in Table~2.

Takeda (1994) has commented that strong Mg lines may not be good 
atmospheric parameter indicators if there is a contribution from an 
extended atmosphere, which would affect the line symmetry.
There is no evidence for line asymmetries in the weak Mg lines used
(here and for the Galactic stars).
Furthermore, these lines have  
high excitation potentials, thus forming deep in the photosphere, 
where any wind or chromospheric component would be negligible.

Microturbulence values in Table~1 have been determined from ionized 
species of Fe, 
Ti and Cr, when enough lines over a variety of line strengths are present.
There is some evidence for slightly different values for each species, 
particularly between neutral and ionized species of the same element 
(e.g., Fe~I and Fe~II).    These differences are typically small
though ($\pm$1-3 \kms), and the weak lines analysed herein are not
very sensitive to the $\xi$ value.   Thus, a single value that best 
fits all species has been adopted per star, with $\Delta\xi$=$\pm$1\kms.
Radial and rotational velocities in Table~1 have been determined 
from spectrum synthesis of Fe~II and O~I lines near 6150~\AA; 
$\Delta$v$_{\rm rad}$ and $\Delta\,v$sin$i$=$\pm$5\kms.

\subsection {Spectral Types and Atmospheric Parameters} 

Examinations of blue supergiants in the Magellanic Clouds have shown 
that the spectral types assigned to these stars are often incorrect 
(Fitzpatrick 1991, Lennon 1997).   This is because the classical
MK classification system cannot be easily applied to metal-poor stars.
A typical mistake is to give a metal-poor B-type star 
a spectral type that is too hot because the metal lines 
are weaker in both cases.
In response to this,
Fitzpatrick (1991) presented a new classification scheme for LMC 
B-type supergiants, and Lennon (1997) has done the same for 
the SMC.   It is striking to see the differences in stars of the same
spectral type due to metallicity alone (compare Lennon's 
Fig.~5 of Galactic stars to his Fig.~6 for SMC stars).

The same is found in this analysis of SMC A-supergiants.
Almost all of the A-type supergiants are cooler than their 
previously assigned spectral types would indicate; see Table~1. 
Some stars are also less luminous than had been previously
indicated from their luminosity classes.  This is an important 
point since some A-supergiants in the SMC were reported to have 
``anomalous'' spectral features such as Balmer lines that are 
too strong for their luminosity classes and the UBV colors that 
are too blue (Humphreys 1983, Humphreys \etal 1991).   It had been
proposed that these anomalies were due to an enrichment of helium
in the stellar atmospheres, however the anomalies are eliminated 
once the cooler temperatures and lower luminosities determined
here are adopted.
All of the ``anomalous'' A-supergiants in this analysis 
have cooler temperatures, e.g., AV~254 and AV~478 were reported 
as anomalous A3~Ia stars (Humphreys 1983, Humphreys \etal 1991), 
whereas the atmospheric parameters are like normal A7~Iab
stars.   Oddly, a nearly identical A7~Iab star, AV~442, 
was classified as a normal A3~Ia star (Ardeberg \& Maurice 1977).   
The H$\gamma$ spectra of these 
three stars are shown together in Fig.~5, which clearly shows they
are nearly identical. 
Ironically, the ``anomalous A3~Ib'' star AV~392 appears to 
be the only properly classified star, A3~Ib, except that it is normal. 

Star AV~213 was also clearly misclassified, or it has changed its
spectral type (an LBV star?).    This star is definitely not a normal 
A2~Iab star (Azzopardi \& Vigneau 1982, Humphreys 1983) as the 
H$\alpha$ profile is in emission (unlike the other supergiants) 
and the metal lines are stronger than most of the other stars.
The atmospheric parameters are typical of a high luminousity
F0~Ia hypergiant.   
In addition, the metal abundances of this star are consistent
to within 0.2~dex, just like the other A-supergiants analysed
which implies the atmospheric analysis is rather robust.

Finally, the atmospheric parameters for two stars, AV~110 
and AV~174 (the hottest and coolest stars in the sample), 
are slightly more uncertain than most since only one 
line of Mg~I and/or Mg~II was available.

\section{ Elemental Abundances}

The chemical abundances in each star have been calculated assuming
the best available atomic data in the literature (see the discussions
in Venn 1995a,b, and references in Appendix~A).   Individual stellar
abundances and estimated uncertainties are listed in Tables~3--6 as
12+log(X/H) $\pm\sigma$ (\#), where \# is the number of lines used
in the average (``S'' in the LTE O~I row refers to spectrum synthesis). 
In three cases, N, O, and Mg, both LTE (italicized) and NLTE corrected 
abundances are listed;  the LTE abundances are only shown for comparison.
Note that the NLTE corrections are calculated for the line formation of
those three individual elements only within an LTE ATLAS9 model 
atmosphere (discussed previously as the best models for A-type supergiants
in Section 3).   Individual spectral line data is listed in Appendix~A.
 
Mean abundances for the sample are listed in Table~7, including the
Galactic A-supergiant results and solar abundances 
(from Anders \& Grevesse, 1989, and Grevesse \& Noels, 1993) for comparison. 
These abundances are also graphed in Fig.~6 as unweighted averages of all 
the stars analysed relative to Galactic A-supergiant abundances.    
Most elemental abundances determined previously for Galactic A-type
supergiants are similar to solar (or to the solar neighborhood in the
case of CNO), a result which was used to emphasize the success of the 
tailored atmospheric analysis of these otherwise complex stars.   

{\it Carbon upper limits:} 
Upper limits can be placed on the C abundances, through 
non-detections of the most accurate C~I indicator, 
i.e., the $\lambda\lambda$7115 multiplet.    
Adopting 15~m\AA\ as a limiting equivalent width 
(which is approximately 1-$\sigma$ in the S/N of the spectra) 
then the upper limits for the cooler stars are intriguing,  
[C/H] $\le -$1.3 for AV~174.  This is consistent with
the value for C derived from the H~II region observations
(see Section 6.1).  
NLTE corrections could lower the C limits further 
(by $\le$0.3~dex, based on calculations using the 
St\"urenburg \& Holweger 1990 C~I/II model atom).
These corrections put the upper limits for the six 
coolest stars in this analysis in agreement with the 
nebular/B-star abundances.

{\it Nitrogen:}  Nitrogen abundances have been determined
from the spectral lines of two N~I multiplets, 
$\lambda\lambda$7440 and $\lambda\lambda$8210.   Sample
spectra are shown in Fig.~7.
Both multiplets include lines of various strengths that are
uncontaminated by other elemental features (except telluric 
lines which are removed by a hot standard star observed near 
the same airmass).    The LTE nitrogen abundances are 
near solar, with only a few exceptions (see Tables 3 and 5),
although NLTE effects are significant.  

In the Galaxy, NLTE corrections range from $-$1.0~dex in 
A0 supergiants to $-$0.3~dex by the F0.  NLTE corrections 
also remove a trend in N with effective temperature and
line strength. 
Using the same N~I/II model atom and techniques as for 
the Galactic stars, NLTE corrections have been computed for 
the nitrogen abundances in the SMC stars.    
The lower metallicity of these stars has a negligible effect
on the statistical equilibrium of N~I relative to their
Galactic counterparts.   As discussed by Lemke \& Venn 
(1996), this is because the statistical equilibrium is 
almost completely controlled by the UV flux around 
1200~\AA\ through the photoionization of the
N~I 2$p^{\rm 2}$P$^{\rm 0}$ ground state.   This spectral
region is dominated primarily by C~I bound-free transitions
and H~I Ly$\alpha$ absorption.   Therefore, uncertainties
in the  NLTE corrections for N~I in the SMC stars are 
similar to those reported by Venn (1995a, Table~7 therein)
for Galactic stars, i.e., typically $\le\pm$0.1~dex. 

The only remaining tricky part for the N~I NLTE calculations 
for the SMC stars is adopting a carbon abundance. 
Since C could not be determined for each star 
in this analysis, log(C/H)+12=7.3 was adopted for the calculations.
This value is indicated by the nebular abundances, it 
is supported by the main-sequence SMC B-stars, and corresponds
to the upper limit on the A-supergiant, AV~174, in this analysis.
If the carbon abundance is reduced to log(C/H)+12=7.0, 
e.g., if mixing of CN-cycled gas has occured which would reduce
C at the stellar surface, then the NLTE N~I corrections are 
not strongly affected (the corrections are larger by $\le$0.03~dex).
If C is higher, e.g., log(C/H)+12=7.6, then the NLTE corrections can 
be slightly smaller, i.e., NLTE N abundances could be $\sim$0.05~dex 
higher. 
 
The mean NLTE nitrogen abundance of 9 A-supergiants in this analysis
is log(N/H)+12=7.33 $\pm$0.35.   The scatter in the abundances is 
quite large; larger than is expected from only atmospheric analysis 
uncertainties.     In particular, in two cases, stars with very 
similar atmospheric parameters have the largest differences in N, 
i.e., N(AV\,213)\,=\,N(AV\,174)\,+\,1.0, 
and N(AV\,478)\,=\,N(AV\,254)\,+0.5.
This alone argues that the star to star variations seen
in the N abundances are real.   In addition, the range in 
the abundances of other elements from star-to-star are quite small 
(see Table~7 and Fig.~6, discussed further in Section~6.2). 

{\it Oxygen:} O~I is determined from spectrum synthesis of the 
$\lambda\lambda$6158 multiplet throughout (see Fig.~8 for 
samples of the spectra), plus the $\lambda$6454 line in AV~442.   
The multiplet comprises three weak lines that are
free of blends in this temperature regime.
Uncertainties in $v$sin$i$ and the S/N of the 
spectra dominate the small LTE oxygen abundance uncertainties,
which are listed in Tables 4 \& 6.
 
NLTE corrections to these oxygen abundances can be estimated
based on the detailed calculations of O~I in A-F stars by 
Baschek \etal (1977) and Takeda (1992).    
Baschek \etal use the complete linearization method and build
a simple 8 level model atom with only 7 explicit line transitions.   
Assuming theoretical photoionization and collisional rates, they 
determine NLTE corrections for various multiplets in modified 
Kurucz (1969) model atmospheres (without line blanketing).
They report a reduction in the theoretical equivalent widths 
for the O~I $\lambda\lambda$6158 multiplet lines by 65\%/15\% for 
models at 10000~K/7500~K with \logg=1.0.  In this analysis,
these translate to abundance corrections of $-$0.5/$-$0.15~dex.
  
On the other hand, Takeda (1992) uses the ALI 
(accelerated-lambda-iteration) method and builds a 
model atom with 86 levels and 294 explicit line transitions.
Adopting different photoionization rates than Baschek \etal
(hydrogenic versus scaled Thomas-Fermi values), and updating
the oscillator strengths, Takeda determines the NLTE corrections 
for various multiplets in newer Kurucz (1979) line blanketed 
model atmospheres (e.g., at 9500~K and 6500~K, with \logg=1.5).    
Remarkably, Takeda's NLTE corrections for the 
$\lambda\lambda$6158 lines are almost identical to
Baschek \etal's.  Oxygen from $\lambda\lambda$6158
in the hotter/cooler supergiants 
is reduced by $-$0.35/$-$0.1~dex.   

In this paper, these NLTE corrections for the oxygen 
abundances in A-F supergiants have been adopted, 
scaling the corrections to reflect temperature and 
luminosity differences, as follows;  

$\Delta$O~I~(\teff$\ge$9000~K)=$-$0.3, 
 
$\Delta$O~I~(8500$>$\teff$\ge$8000~K and \logg$>$1.5)=$-$0.2, 
 
$\Delta$O~I~(8500$>$\teff$\ge$8000~K and \logg$\le$1.5)=$-$0.1, 
 
$\Delta$O~I~(\teff$<$8000~K)=$-$0.1 (see Tables 3 \& 5). 

The mean NLTE corrected oxygen abundance is then 
log(O/H)+12=8.14.  Retaining the uncertainty from the LTE 
spectrum syntheses (which should continue to dominate the 
abundance uncertainties), then [O~I/Fe~II]=0.00~$\pm$0.10.   
Oxygen, however, is higher in the Sun than in other objects
in the solar neighborhood, including Galactic A-supergiants.
Adopting the same NLTE corrections for the Galactic A-supergiant
abundances in Venn (1995a), then NLTE O(GAL AI's)\,=\,8.59.
Thus, the differential abundance is $-$0.45~dex 
(i.e., log(O/H)$_{\rm SMC} -$ log(O/H)$_{\rm GAL~AI}$).

Adopting the same NLTE corrections for the Galactic A-supergiants
assumes that there is no significant metallicity dependence
to them.    This is not a bad assumption considering 
that Takeda and Baschek \etal derived very similar results 
using different initial O abundances (12+log(O/H)=8.8 and 8.6, 
respectively) and different model atmospheres 
(e.g., Takeda's included line blanketing). 

Thus, the differential O underabundance, $-$0.45~dex, 
is slightly smaller than the iron underabundance, $-$0.6~dex,
but it is in good agreement with the 
differential O abundance from nebulae and B-stars,
e.g., $-$0.5~dex (discussed in Section 6.1). 
 
{\it Sodium:} Na has been determined from the Na~D
resonance lines, and/or Na~I (mult. \#4) at $\lambda$8194 in six stars.
Sodium abundances range from being in good agreement with the iron-group
underabundances, to showing slight enhancements.  The average of this
sample is [Na~I/Fe~II]=+0.23 $\pm$0.2.     

A large overabundance of Na was seen in the Galactic A-type supergiants 
from the same spectral features, [Na/H]=+0.66.  Overabundances of Na 
have also been found in Galactic F-K supergiants and interpreted in 
a number of ways (see the review by Lambert 1992).  NLTE effects have 
been ruled out by Drake \& Lambert (1994), 
but other failures of the model atmospheres 
analyses may still exist.   In the SMC, it is also unclear as to what 
the initial Na abundance should be in young stars, making any interpretation 
based on stellar surface changes difficult.

{\it Iron-group:} 
Fe~II, Cr~II, and Ti~II are determined from several lines 
throughout the optical spectra, and from a variety of multiplets 
with different line strengths and excitation potentials.  
The mean of these abundances is [M/H]=$-$0.7 $\pm$0.2 (or $-$0.6~dex
relative to Galactic A-supergiants). 
Uncertainties in the atmospheric parameters have only small effects
on these abundances ($\Delta$(Fe~II, Cr~II)$\le\pm$0.14~dex, 
$\Delta$(Ti~II)$\le\pm$0.2~dex). 

Fe~I is measured from many weak lines, and is typically 0.0 to 0.3~dex 
lower than Fe~II.    This is consistent with the underabundances 
predicted from NLTE calculations in Galactic A-dwarfs (Gigas 1986) and 
early-F supergiants (Boyarchuk \etal 1985).   
NLTE effects cause Fe~I to be overionized by the UV radiation field, 
which one might expect to be affected by metallicity through line blanketing, 
although Gigas (1986) examined metallicity effects  in Vega
([Fe/H]=0 and $-$0.5), and found negligible differences in the NLTE 
corrections.   

Sc~II is consistent with the iron-group results, although measured 
from far fewer lines.   Additionally, Cr~I is determined in three stars
(AV~463, 213, and 174, from 2-3 lines of multiplet \#7 near $\lambda$5206), 
and is in good agreement with Cr~II throughout.

{\it Alpha-elements:}  Mg, Si, and Ca (Ti is discussed above with the
iron-group elements) are primarily synthesized together during 
explosive O and Si burning through the addition of $\alpha$ particles 
to lighter seed nuclei.   Most of the $\alpha$-element underabundances 
determined here are in excellent agreement with the iron-group 
underabundances.   

Magnesium, determined from weak and unblended Mg~I and Mg~II lines
(that have been critically tested for accuracy, see Venn 1995b) 
is always consistent with the iron-group underabundance 
(to within 1$\sigma$).   
Uncertainties in the atmospheric parameters hardly affect Mg~II,
making it an excellent abundance indicator in this temperature 
regime.

Silicon abundances are measured from 2 to 6 lines of Si~II in all 
the target stars.   In most stars, the silicon depletion is the 
same as the iron-group and Mg depletions to within 0.2~dex.    
Some of the hotter stars appear to be slightly overabundant,
[Si/Fe]$\ge$0.3, which may suggest an increasing NLTE effect.
This is {\it not} similar to the results from Galactic A-supergiants, 
where the mean was [Si~II/H]=$-$0.2 $\pm$0.2.   
Only one line of Si~II was observed in only five Galactic stars,  
but additional Si~I abundances were computed in 11 of the cooler 
Galactic A-supergiants from two lines 
that yielded [Si~I/H]=$-$0.1 $\pm$0.1.   
It is difficult to understand this trend in the hotter SMC stars, since 
the analysis of Si~II is expected to yield reliable results, although 
Si~III is the dominant ionization stage in these stars meaning that 
metallicity-dependent NLTE effects might be present.

Derived underabundances of Ca~II are consistent with the iron-group 
underabundances throughout.   Ca~I is also found for the five coolest 
stars (primarily from the $\lambda$4226 line of multiplet \#2), 
and is in fair agreement with Ca~II.

{\it s-process elements:} Zr~II and Ba~II have been determined 
from one or two lines each per star (see Appendix A), and are similar to the 
iron-group underabundances to within 0.3~dex.  This may or may not be
a significant difference considering how few lines have been analysed.
 
Sr~II appears more underabundant, with [Sr/Fe] $\le$1.0~dex, but
NLTE effects are expected for the two Sr~II resonance lines 
used here.   Lyubimkov \& Boyarchuk (1982) found that Sr abundances 
measured from the resonance lines are 0.2 to 0.4~dex less than those 
from a subordinate line in Canopus (F-supergiant) 
due to NLTE effects.  Praderie (1975) found
theoretical equivalent widths may be two times smaller 
in main-sequence A-stars, also due to NLTE effects. 
The primary Ba indicator in this analysis is also a resonance line.
Lyubimkov \& Boyarchuk found the same effect for the Ba~II 
resonance lines as for Sr~II, however in one star, AV~174, 
analysed here, the same abundance is determined 
from a subordinate Ba~II line as from the resonance line.
Therefore, NLTE effects on the Sr and
Ba lines remain uncertain.

\section{Comparisons with SMC F-K supergiants } 

The chemical composition of SMC F-K supergiants has been determined
by several authors (Russell \& Bessell 1989 = RB89, 
Spite \etal 1989a,b = S89, Luck \& Lambert 1992 = LL92, 
Hill \etal 1997a,b = H97, Luck \etal 1998 = L98). 
These results are shown in Table~8 as differential abundances,
i.e., the abundances listed are the differences between the
average for the SMC stars and that of Galactic standard stars
as analysed by each individual analysis (see the footnotes to
the Table for the standard stars used from each analysis).   
These differential abundances are a much more accurate and
useful way to compare the results from so many different analyses 
since differential results can minimize the systematic errors per analysis.
Differential abundances were first recognized as an improved way 
to compare abundance data from different Magellanic Cloud objects
by Pagel (1992, ``like-like'' comparisons).

{\it Iron-group:} 
The most striking result is that the Fe abundances are essentially 
identical, [Fe~II/H]$\sim -$0.6.   A histogram of Fe determinations 
from A- through K-type supergiants shows that the scatter within
the analyses and between the analyses is very small (Fig.~9). 
Previously, stars in NGC~330 were thought to be more metal-poor than
field stars in the SMC (e.g., Spite \etal 1991, Grebel \& Richtler 1992),
but recent Fe abundance determinations in 
F-K supergiants by H97 and Si abundances in B-giants 
by Korn \& Wolf (1998) show no significant differences between the
cluster and field stars.
Other iron-group elements Sc, Ti, Cr (and Ni in F-K supergiants) 
also show similar underabundances to Fe ($\pm$0.2~dex) in all analyses.

{\it Carbon:} 
The differential results (Galactic minus SMC) from F-K supergiants 
suggest that C has roughly the same underabundance as iron, 
e.g. $-$0.6 to $-$0.8~dex (see Table~8). 
This may be slightly inconsistent with the nebular and 
B-star abundances which show slightly larger C underabundances
(discussed in Section 6.1), e.g., the nebular C abundance 
is about 0.9~dex less than in Orion 
(and Orion is about 0.3~dex less than solar, 
c.f., Cunha \& Lambert 1994, 
which conspires to yield SMC [C/H]=$-$1.2).   

The upper limit on carbon from AV~174, the coolest
star in this analysis, is in better agreement with the 
lower nebular abundance than the slightly larger F-K 
supergiant abundance.
If mixing of CNO-cycled gas has occured in AV~174, then 
this could have lowered C from an initially higher abundance
(such as that reported from F-K supergiant analyses), 
but it seems unlikely that mixing has occured in this 
star since N is lowest of all stars examined (see Table~5).  
 
It is unlikely that the F-K supergiant C abundances are in 
error since several different features of C have been 
used in the most recent abundance determinations;   
L98 examined both C$_{\rm 2}$ molecular features and C~I
absorption lines, H97 analysed $^{\rm 12}$CN, 
$^{\rm13}$CN, and C$_{\rm 2}$ molecular features, LL92
measured permitted and (one) forbidden C~I line, 
and S89 and RB89 used C~I lines.  
In the past, the RB89 and S89 supergiant C results 
were used as an example of the inconsistencies between stellar
and nebular abundances in the Clouds, yet those abundances
were much higher still (see Table~8).   The recent abundances
are much more accurate, e.g., RB89 determined C from only one 
feature in only 2 stars.  
 
{\it Nitrogen:}
Nitrogen abundances in SMC supergiants have a very large range, 
[N/Fe]$>$1.0, but this range is consistent between analyses, i.e.,
all analyses that include several stars find a wide range in N
(see the histrogram in Fig.~10).    These N abundances have been
determined from a variety of features, including N~I lines (this paper,
LL92), N~II lines (R93, Lennon \etal 1996), and molecular 
CN features (H97, Barbuy \etal 1991).   
The range itself is not due to uncertainties in the individual analyses, 
nor due to differences between the analyses (although a small NLTE
correction required for early F supergiants can account for the slight 
offset in the LL92 data due to two stars with N$>$8.0 in Fig.~10).  
Thus, the range in N is astrophysically significant and 
is discussed below in the context of stellar evolution (Section~6.2).

{\it Alpha-elements:} 
The $\alpha$-elements typically have the same underabundance 
as [Fe/H] ($\pm$0.2 dex), a result that is consistent between 
the (differential) results of all of the stellar analyses listed 
in Table~8, with only a few exceptions. 
Minor exceptions seen in only one of the analyses can usually
be attributed to uncertainties within that analysis, 
e.g., RB89's high Ca~II abundance was calculated from 
only one line in one star.
Silicon, on the other hand, appears to be slightly overabundant 
(with respect to iron) in some of the supergiant analyses.  
Slight silicon overabundances are probably due to uncertainties 
in the atmospheric analyses, as in the case of the A-supergiants 
(discussed in Section 5, 
where the overabundances appear related to temperature). 
In addition, most other supergiant analyses have [Si/Fe]=0,
other $\alpha$-elements do not show an overabundance, 
and the B-stars analysed by Rolleston \etal (1993) 
have [Si/Fe]=0.

Especially noteworthy is the very good agreement in the O
and Fe abundances between the supergiant analyses listed in
Table~8, and thus the excellent agreement in the O/Fe ratios.
Additionally, the O abundance determined from supergiants
is in excellent agreement with that determined from B stars
and H~II regions (discussed further below).   Thus, O/Fe
is like the other $\alpha$/Fe ratios in that O and $\alpha$ 
elements have the same underabundance as Fe; however it is
necessary to report [O/Fe]=$-$0.2 because the Sun is slightly 
O rich relative to the solar neighborhood (discussed further
below). 

{\it s-process elements:}
Good agreement is found between the three s-process elements
examined here in the A-type supergiants (Sr, Zr, and Ba) and
F-K supergiant results.  These three elements share the 
underabundances of the iron-group elements, and are often
slightly lower.  Sr by RB89 and herein should not be weighed
too heavily since RB89 measured only one (strong, 394 m\AA) line
in only one star, and the A-supergiant Sr suffers from neglected
NLTE line formation effects.

\section{Stellar Evolution}

Elemental abundances can be used to examine the evolutionary
status of a star since hydrogen-burning occuring in interior
layers may be mixed with pristine gas at the surface of a star
during its evolution.   Since massive stars burn hydrogen 
primarily via the CNO-cycle, then the ratios of these catalysts
change in a predictable way.    
Although many evolution scenarios exist, few are well constrained
by the observations for massive stars, particularly for low
metallicities (c.f., Langer \& Maeder 1995, Maeder \& Conti 1994,
Stothers \& Chin 1992).
For example, the conditions as to when massive stars burn helium 
as blue or red supergiants remain uncertain, and the evidence of 
CN-cycled gas at the surfaces of main-sequence B-stars (which are 
expected to have pristine, unmixed, abundances) is puzzling.   
  
Most stellar evolution scenarios at low metallicities predict
that massive stars burn helium in their cores as blue supergiants.
As an example, Schaller \etal (1992) predict that stars with 
M$<$20 M$_\odot$ burn helium while evolving along a blue loop, 
while stars $\ge$20 M$_\odot$ burn helium while evolving directly 
towards the red giant branch.   In this scenario, the most 
massive blue supergiants would have unaltered CNO abundances,
while the less massive blue supergiants 
would have first dredge up abundances 
(where the change in the 12+log(X/H) abundances
are predicted as C: N: O $\sim$ $-$0.2: +0.5: $-$0.05).
This is a testable prediction.
 
The assumptions required for massive star evolution models
are poorly constrained though, such that blue loops may not 
exist at all, or perhaps all stars burn helium as blue supergiants
at low metallicities (as either post- or pre- red supergiants).   
Langer \& Maeder (1995, their Fig.~1) show an additional complication; 
in their investigation of semiconvection at z=0.002, 
they predicted three different 
evolutionary scenarios for three different masses, 15, 20, and 
25 M$_\odot$, in otherwise identical models.  Helium burning occurs 
during a blue-loop at 15 M$_\odot$, during the red supergiant phase 
at 20 M$_\odot$, and as a post-main-sequence blue supergiant at 
25 M$_\odot$.  Clearly, reliable abundances of CNO for blue supergiants 
in the Magellanic Clouds will help to constrain these models.

Additionally, the effects of rapid rotation on massive stars have 
been investigated theoretically for Galactic stars (Maeder 1987, 
Dennisenkov 1994, Fliegner \etal 1996, Meynet 1997, Maeder \& Zahn 1997,
Talon \etal 1997, Langer \& Heger 1998), as well as LMC stars
(Langer 1991).  
With respect to abundances, star-to-star variations
of light elements in some Galactic OB-stars and blue supergiants 
require a more discrete parameter like rotation rate.    A few
main-sequence B-stars were reported to have enhanced N 
(Gies \& Lambert 1992), while several B and A-supergiants are
also slightly N enhanced (Venn 1995a, Lennon 1994). 
In addition, boron is expected
to be destroyed by partial mixing on the main-sequence so that
a B-N trend should exist, although at present very few boron abundances
are available for an examination (c.f., Venn \etal 1996, Cunha \etal 1997). 
As rotation rate is a metallicity-{\it independent} parameter, then partial
mixing may also affect the Magellanic Cloud stars.

\subsection {Initial CNO Abundances} 

To interpret the CNO abundances in Magellanic Cloud stars in the
context of stellar evolution, it is necessary that we know initial
as well as current abundances to look for signs of mixing.   The best
way to estimate the initial abundances of young stars is by examining 
CNO from nebular abundances and from main-sequence B-stars.    
Since CNO each have different nucleosynthetic sites from each other 
and from Fe-group elements, it is {\it not} sufficient to assume that they
scale with the overall metal underabundance of the SMC (as has often
been assumed in previous analyses).

Nebular abundances in the Magellanic Clouds have been determined
by several investigators (c.f., Dufour 1984, 1990; Pagel 1992).   
More recent SMC nebular analyses done with modern detectors have
been reported by Russell \& Dopita (1990), Garnett \etal (1995),
and Kurt \etal (1998). 
These abundances are discussed further below.
On the other hand, only a few calculations of abundances in 
unevolved B-stars in the SMC exist (Rolleston \etal 1993 = R93), 
but these should reflect those of the interstellar medium.

For carbon, absolute abundances from B-dwarfs in the SMC are very 
uncertain due to NLTE effects of the C~II feature analysed.   
For example, R93 report an {\it absolute} abundance for carbon of
12+log(C/H)=6.9, significantly lower than the nebular abundance; 
but, taking the mean of the Galactic B-stars and subtracting the
{\it differential} abundance for the SMC B-stars yields, 
12+log(C/H)=8.3 $-$0.9 =7.4, in perfect agreement with the
nebular abundances (see Table~10).  R93 also comment that this
is the correct way to interpret their C abundance because NLTE 
is expected to strongly affect their C~II $\lambda$4267 feature.
The fact that the SMC B-star C abundance agrees well with the 
SMC nebular result is significant since it implies that depletions 
of C onto dust grains in the ISM are small!   
This result has also been found for nebular and B star C 
abundances in Orion.

Mathis (1996) reviews and discusses the observations 
of CNO in the interstellar medium in the Galaxy.   Very little N is 
expected to be in dust, however only 2/3 of the O atoms and half of 
the C atoms are thought to be in the gas-phase (also supported by 
more recent observations by Meyer \etal 1998 and Sofia \etal 1997).  
Thus, the nebular abundance of N should be the same as young B-stars, 
but nebular C and O should be slightly less than the B star abundances, 
by roughly 0.15~dex and 0.3~dex, respectively. 
Such large differences between nebular and B star 
abundances are not seen though; for example, the nebular and
B star abundances in Orion are in excellent agreement 
(see Table 9; also discussed by Mathis 1996).   
Recently, Esteban \etal (1998) adopted slightly smaller 
dust correction factors for C and O, $\le$0.1~dex
for their Orion gas abundances.  These smaller corrections 
are well within the 1 to 2$\sigma$ uncertainties in
the stellar abundances.  

Assuming that global metallicity does not change the element depletion 
fractions of the gas-phase ISM too significantly, then the nebular and 
B star CNO abundances should be in good agreement in the SMC as well. 
Thus, the agreement between the nebular and B stars
C abundance strongly suggests that 12+log(C/H)=7.4 is the correct 
present-day C abundance, as opposed to the higher C abundance 
reported from the F-K supergiants.   
Ironically, the difference between the F-K supergiants
and the nebular/B-stars C abundance is about +0.3~dex, which is
the predicted interstellar dust depletion factor for the ISM.  
For this paper, the present-day C abundance is not critical since upper 
limits only have been determined for C in A-type supergiants, however
the present-day N abundance is critical. 
 
The SMC nebular nitrogen abundances are very low, 12+log(N/H)=6.6;
this has been determined by Dufour (1984) and RB89 assuming 
N+/O+ = N/O, which Garnett (1990) showed is sufficient for an accurate 
N abundance in low metallicity galaxies from photoionization model tests. 
As further confirmation, recent N++/O++ measurements (Garnett \etal 1995,
Kurt \etal 1998) of one high excitation nebula, SMC N88A, 
has resulted in a similar low N abundance, see Table~10.   
In addition, grain depletion of N is expected to
be negligible (see above), and since N+/O+ and N++/O++ yield such 
similiar results, then this implies that the ionization corrections
are well established.
In the B-stars, only an upper limit on the 
N abundance is available because no N~II lines were observed (a marginal
detection of a single N~III line in one star yields consistent results
but is uncertain due to a blend with an O~II line). 
Thus, the nebular N abundance is adopted in this paper as the 
initial abundance for the supergiants.  

The abundance for oxygen determined from the SMC B-stars is in
excellent agreement with that from the SMC nebulae (both absolute 
and differential O abundances are in agreement), 12+log(O/H)=8.1,
as seen in Table~10.   This value is also found from analyses of
evolved B-stars (Reitermann \etal 1990, J\"uttner \etal 1992,
Rolleston \etal 1993, Lennon \etal 1996, Korn \& Wolf 1998) and
is consistent with the differential results for F-K supergiants 
(see Table~8).

\subsection {Implications for Massive Star Evolution}

For Galactic stars, examination of CNO in supergiants is not too 
difficult since detailed abundance analyses of CNO have been done 
for over 70 B-stars in the solar neighborhood and nebular abundances 
are available for Orion and other H\,II regions throughout the Galactic 
disk (see Table~9 for references and results).   From these analyses, 
evidence for slight mixing of CN-cycled gas has been found in some 
main-sequence B-stars and B-giants (Gies \& Lambert 1992, Lennon 1994), 
as well as in the A-type supergiants analysed by Venn (1995a,b).    
This can be seen in a histrogram of the N abundances (Fig.~10), 
by the slight high-end tail in the Galactic B-stars, and the offset 
from the initial N abundance in the Galactic A-type supergiants.
In the SMC, only 3 main-sequence B-stars have been examined 
in detail, but there are more data from nebular abundances (see Table~10).
The initial CNO abundances adopted here are the best available, however
they remain a major source of uncertainty, especially for C which differs
by 0.3~dex from F-K supergiant analyses (see discussion above in Section 6.1).  
An effort for more abundance analyses of main-sequence 
B-stars in the Magellanic Clouds 
should be undertaken when larger telescopes and/or higher efficiency 
spectrographs are available at the southern observatories.

To examine the evolutionary status of the SMC A-supergiants, we need to 
examine the N abundances for signs of mixing of CN-cycled gas.   Normally,
one likes to examine the N/C ratio since both C and N are affected by
mixing (and the sum of the catalysts, C+N+O, remains constant), but only 
upper limits were obtained for C, and nevertheless N is the element 
most affected.    
The majority of A-supergiants in the SMC show N 
enrichments over the nebular/B-star abundance.    The amount of the
enrichment varies from $\le$0.2~dex to $\ge$1.2~dex!
This range is large, and suggests that enrichments 
can vary drastically from star-to-star.   
A similar phenomenon is seen in the SMC B-giants, since 
Lennon \etal (1996) found that two B-giants in NGC~330 appear 
to be very enriched in N, yet two other B-giants in the same 
cluster analysed by J\"uttner \etal (1992) report the opposite.   
 
The range, from close to the initial N value to well beyond the 
first dredge-up value, cannot be due to atmospheric uncertainties, 
e.g., Fig.~9 shows the range in N is much larger than for O or Fe.
In addition, the median value is greater than the predicted first 
dredge-up abundance (of $\sim$+0.6~dex).  Perhaps all of these 
stars are post-red supergiants, however this ignores the fact 
that some stars appear to have undergone little or no mixing.    
If most stars have undergone the first dredge-up,
and only a few are post-main-sequence stars, 
then this could account for the lower N values 
(standard models predict only 1 in 50$-$100 stars in this mass range
would be post-main sequence stars, yet some models, 
e.g., Langer \& Maeder, 1995, predict closer to even numbers); 
but the highest N abundances still remain unexplained. 
 
The fact that partial mixing is indicated in the Galactic stars
affects the interpretation of the SMC abundances --   
perhaps most SMC stars undergo partial mixing,
in addition to the first dredge-up.
If so, this could account for the highest N abundances
(first dredge-up and extreme rotational mixing), as well as the 
lowest results (slight rotational mixing only).   Of course, this 
implies a mixed population of A-type supergiants, including both 
post-RGB and post-MS stars, but this is already predicted by 
many evolution scenarios 
(e.g., Schaller \etal 1992, Bertelli \etal 1994, Langer \& Maeder 1995).

Now the question becomes ``could rotational mixing account for the
N abundances entirely?''    At present, only the most rapid rotators 
are expected to undergo significant main-sequence mixing that would 
enrich N as efficiently as the first dredge-up.   Therefore, if 
rotational mixing is entirely responsible, then main-sequence mixing 
must be more efficient in metal-poor stars.   This means the stars 
must all rotate rapidly.   Although data on rotation rates in SMC 
main-sequence B-stars is scant, this seems rather unlikely.    

Although carbon abundances could not be determined here as a further
test of this interpretation, there is one other element that may be an
indicator of mixing -- sodium.     
In this paper, A-supergiant [Na/Fe] is near solar, which is similar
to the recent results by L98 for F-K supergiants. 
L98  interpret their {\it lack} of an overabundance as evidence 
{\it against} mixing processes, assuming that
enrichments of Na occur due to mixing of Ne-Na cycled gas. 
This interpretation is complicated though, 
in part because several different explanations have been proposed 
to explain Na enrichments in Galactic stars (see Lambert 1992), but 
more importantly because we do not know the initial Na abundance in 
the SMC.
  
One way to use Na though is to look for a Na-N or Na-O trend.  
If Ne-Na cycled gas has mixed to the surface, then CN-cycle/ON-cycled 
gas would accompany it since those cycles run at cooler/similar temperatures. 
Any mixing of primary Na must also enrich secondary N (and deplete C and O).
Plots of Na versus N and O are shown in Fig.~11.   It might
be possible to see some hint of a trend in the data 
(e.g., Na increasing with N, but decreasing with O), 
but nothing significant because of the large scatter in the
data.   Theoretical first dredge-up values (for classical 
non-rotating stellar evolution models by Heger 1998) are 
shown in Fig.~11 by arrows (adopting initial abundances 
for N and O from Table~10, and assuming [Na/Fe]=0).

\section {Chemical Evolution of the SMC}

Several chemical evolution models exist for the SMC, in which 
various assumptions are invoked (e.g., inflow of unprocessed gas, 
Russell \& Doptia 1992, steeper IMF, Tsujimoto \etal 1995)  
to reproduce the low $\alpha$ element abundances determined 
from stars, planetary nebulae, supernovae, and H~II regions, 
relative to the solar neighborhood.  
The ratio of, for example, O/Fe is an important constraint since
these elements have different nucleosynthetic sites; oxygen is 
primarily produced in Type II supernovae events, while iron comes 
from all supernovae, including SN~Ia during
quiescent intervals (Gilmore \& Wyse 1991). 
More recently, Pagel \& Tautvai\u{s}ien\.e (1998) have suggested
that large variations on the Galactic chemical evolution models 
are not necessary to reproduce the abundances, partially because 
the $\alpha$-element abundances are not unusually low when considered 
in ``like-like'' analyses with their Galactic counterparts.     
Pagel \& Tautvai\u{s}ien\.e assume the same yields and time delays 
as for their solar neighborhood chemical evolution models to produce 
analytical chemical evolution models for the Magellanic Clouds.

It is certainly true that oxygen is not unusually low when 
examined relative to Orion or the solar neighborhood, and not to the Sun.
When oxygen is examined this way, then the SMC underabundance 
of O is similar to that of Fe (discussed in Section 5).   
For example, 12+log(O/H)=8.9 in the Sun, but only 12+log(O/H)=8.6 from
B-stars in the solar neighborhood, B-stars in Orion, and nebular Orion
abundances (see Table~9).  
12+log(O/H)=8.6 is also the average determined from 
analyses of young F-K supergiants (e.g., Luck \& Lambert 1985, 
Barbuy \etal 1996, L98), A-type supergiants (Venn 1995a,
with NLTE corrections sited in Section~4), and some B-type supergiants 
(e.g., Lennon \etal 1991, Gies \& Lambert 1992) in the Galaxy.
Oxygen abundances in the SMC are $-$0.5~dex from this value, as
determined by the B-stars and nebulae (see Table~10), which is
supported by the abundances found from SMC supergiants (see Table~8).   
Thus, [O/Fe]=$-$0.2 (recalling that the square bracket notation 
denotes an abundance relative to the Sun), but more accurately the 
differential abundances are log(O/Fe)$_{\rm smc} -$ log(O/Fe)$_{\rm gal}$=+0.1. 

A similar result is found for the other $\alpha$-elements.
In this analysis, the abundance of Mg is particularily well 
determined --  Mg~II is almost insensitive to uncertainties 
in the stellar parameters (see Tables~4 \& 6), yielding a 
very reliable abundance.  [Mg/Fe]=0 in the SMC 
A-supergiants (see Table~7).  Similar underabundances have 
been found from other analyses of SMC F-K supergiants, 
and B-stars (see Table~8), but from fewer lines that are
more sensitive to atmospheric uncertainties in both cases.

Other $\alpha$-elements determined here include Si, Ca and Ti
(which acts like an $\alpha$-element in Galactic stars; c.f.,
Edvardsson \etal 1993, and the discussion by Wheeler \etal 1989).
Si~II abundances in the A-type supergiants appear to be
uncertain in some supergiant analyses (discussed in Section~4).   
The Ca abundances from Galactic A-supergiants are not consistent, 
with [Ca~I/H]=+0.3 and [Ca~II/H]=$-$0.3.   Examining only 
the Ca~II abundances, which should be better determined in these 
stars since Ca~II dominates the ionization balance over Ca~I
(where NLTE effects may be strong), then the
like-like depletion of Ca in the SMC is $\sim$0.7~dex,  in
excellent agreement with O and Mg.   This abundance is also
in good agreement with the Ca~I abundances determined in F-K
supergiants (see Table~8). 
Ti~II is well determined in both Galactic and SMC A-supergiants.
Even though the abundances are rather sensitive to the
atmospheric parameters, there are many absorption lines  
available over a range of wavelengths, line strengths, and
formation depths.  [Ti/H]=$-$0.6~dex in the SMC A-supergiants, 
in agreement with the other $\alpha$-elements.   The reliable
Ti~II abundances determined in A-supergiants are also an 
excellent complement to, and in good agreement with, the reliable 
Ti~I abundances determined in the F-K supergiants. 

For comparison to the Galaxy, we have the incredible efforts
by Edvardsson \etal (1993) who have determined abundances in 189 
field F-G disk dwarf stars.  Metal abundances range from 
$-$0.8 $\le$ [Fe/H] $\le$ +0.2, with an estimated accuracy 
of $\pm$0.05~dex.    Their analyses find that O, Mg, and other
$\alpha$-elements do not track the Fe depletions, but instead 
are slightly less depleted than Fe at low metallicities.
At the SMC metallicity of [Fe/H]$\sim -$0.6, 
then [O, Mg/Fe]$\sim$+0.2 to +0.3, and [Si, Ca, Ti/Fe]$\sim$+0.15.
In addition,  Tomkin \etal (1995) found [C/Fe]$\sim$+0.3.
The Galactic overabundances of $\alpha$-elements at low Fe
are not seen in the SMC.    The young stars and nebulae in 
the SMC show present-day, young Galactic abundance ratios
to within their 1-$\sigma$ uncertainties.
These abundance ratios are reproduced by the latest 
chemical evolution models of 
Pagel \& Tautvai\u{s}ien\.e (1998).   

Finally, the light s-process elements determined here can also
be used to examine some details of the SMC's chemical evolution.   
In particular, Pagel \& Tautvai\u{s}ien\.e (1998) consider 
different star formation rate (SFR) histories -- one which has been
constant over time (``smooth'' model), and another with punctuated
episodes of star formation (``bursting'' model).   The change in
the SFR history can affect the age-metallicity diagram for the SMC
as well as the certain present-day s-process elemental abundances.
In this paper, [Zr \& Ba/Fe]=$-$0.2 to $-$0.4, in agreement with 
most F-K supergiant analyses (see Table~8).   [Sr/Fe]=$-$0.6 is also
in good agreement with results by RB89, yet suffers from 
significant analysis uncertainties (discussed in Section~4).   These 
less-than-solar ratios favor Pagel \& Tautvai\u{s}ien\.e's bursting 
model, i.e., their bursting model predicts [Y, Zr \& Ba/Fe]$\sim-$0.2
at present.
In their smooth SFR models, solar-like ratios are predicted.   
It is not possible to distinguish between their bursting and smooth 
models using any other element ratios (e.g., $\alpha$- or Fe-group elements).
Pagel \& Tautvai\u{s}ien\.e's models also make predictions for other
(heavier) s\&r-process elements not discussed here (i.e., not observed
in A-supergiant spectra).   Abundance predictions for those other elements
(e.g., Ce, Nd, Eu) are in poor agreement with the observed abundances from
F-K supergiant analyses, although the uncertainties in those derived 
abundances are also quite significant (e.g., those abundances are often
based on only a few spectral lines with large line-to-line scatter).
Recently, DaCosta \& Hatzidimitriou (1998) have used abundance determinations 
based on the Ca II triplet in red giants in SMC clusters to suggest
that the simpler, smooth model reproduces the age-metallicity trend in
the SMC.   However, Mighell \etal (1998) have analysed HST color-magnitude
diagrams of the same clusters (mostly) and determined ages, metallicities,
and reddenings that may favor the bursting model -- in agreement with
the Zr and Ba abundances presented here.

\section {Conclusions}

Analysis of 10 A-type supergiants in the SMC has shown that
atmospheric parameters and elemental abundances can be well
determined in a tailored analysis (including ATLAS9 LTE model atmospheres,
weak lines from dominant ionization stages, and NLTE line formation only
when necessary).    The atmospheric parameters
show that previously assigned spectral types are often incorrect,
which had resulted in some stars being erroneously classified 
as helium-rich anomalous stars.
 
New elemental abundances of N, O, Na, Mg, Si, Ca, Sc, Ti, Cr, Fe,
Sr, Zr, and Ba have been determined.
[Fe/H]=$-$0.7 $\pm$0.07, but with respect to Galactic A-type
supergiants this is reduced to $-$0.6~dex.    This result is in 
excellent agreement with those from F-K supergiants, and with
other iron-group elements (Ti, Cr). 
$\alpha$-elements have the same underabundances as iron-group
elements, including Mg, Si, and Ca.   [O/Fe]=$-$0.2, but considering 
that the Sun is overabundant in oxygen by 0.3~dex, then actually O 
has the same underabundance as iron (within 0.1~dex).
This is in agreement with the {\it differential} B~star abundance,
the {\it differential} F-K supergiant results, and the
the nebular analyses. 
Certain s-process elements are more underabundant than iron,
[Sr, Zr, Ba/Fe]$\le$0.3~dex, in good agreement with results from
F-K supergiants.   

The N abundances in the SMC A-supergiants strongly suggest 
that these stars have undergone the first dredge-up, as is
currently predicted by most SMC stellar evolution theories.
The large range in these abundances suggests that there
are additional star-to-star variations though, most likely 
related to effects of rotational mixing on the main-sequence
(not currently included in most stellar evolution theories,
but possibly having wide reaching implications on massive star 
ages and masses).  
In addition, a few stars appear to have very low N 
suggesting that they may be evolving from the main-sequence 
only now and have not yet undergone the first dredge-up.

O, Mg, and other $\alpha$-elements show similar depletions
as Fe relative to Galactic standards.   This is different from
metal-poor stars in the solar neighborhood.   The SMC galaxy 
evolution model by Pagel \& Tautvai\u{s}en\.e (1998) reproduce these
abundances.   In addition, higher depletions of certain s-process
elements (Zr, Ba) favor their model with a bursting star formation rate.  

\begin{acknowledgements}{}
Many thanks to Rolf-Peter Kudritzki for making this project 
possible by giving me the opportunity to work with him at 
the Universit\"at-Sternwarte-M\"unchen \& MPA-Garching.   
I am indebted to Danny Lennon for many valuable conversations
about hot stars and the Magellanic Clouds, to Evan Skillman
and Don Garnett for several discussions on nebular abundances, 
to Michael Lemke for advice and help on NLTE analyses, and
to Norbert Langer for guidance in stellar evolution. 
This project has been supported in part by a Max Planck Institute
post-doctoral fellowship and a Clare Boothe Luce Professorship 
grant.   Travel support by the European Southern Observatory
is gratefully acknowledged, as well as superior observing advice 
and assistance.

\end{acknowledgements}

\vfill\eject

\begin{figure}[t]
\plotfiddle{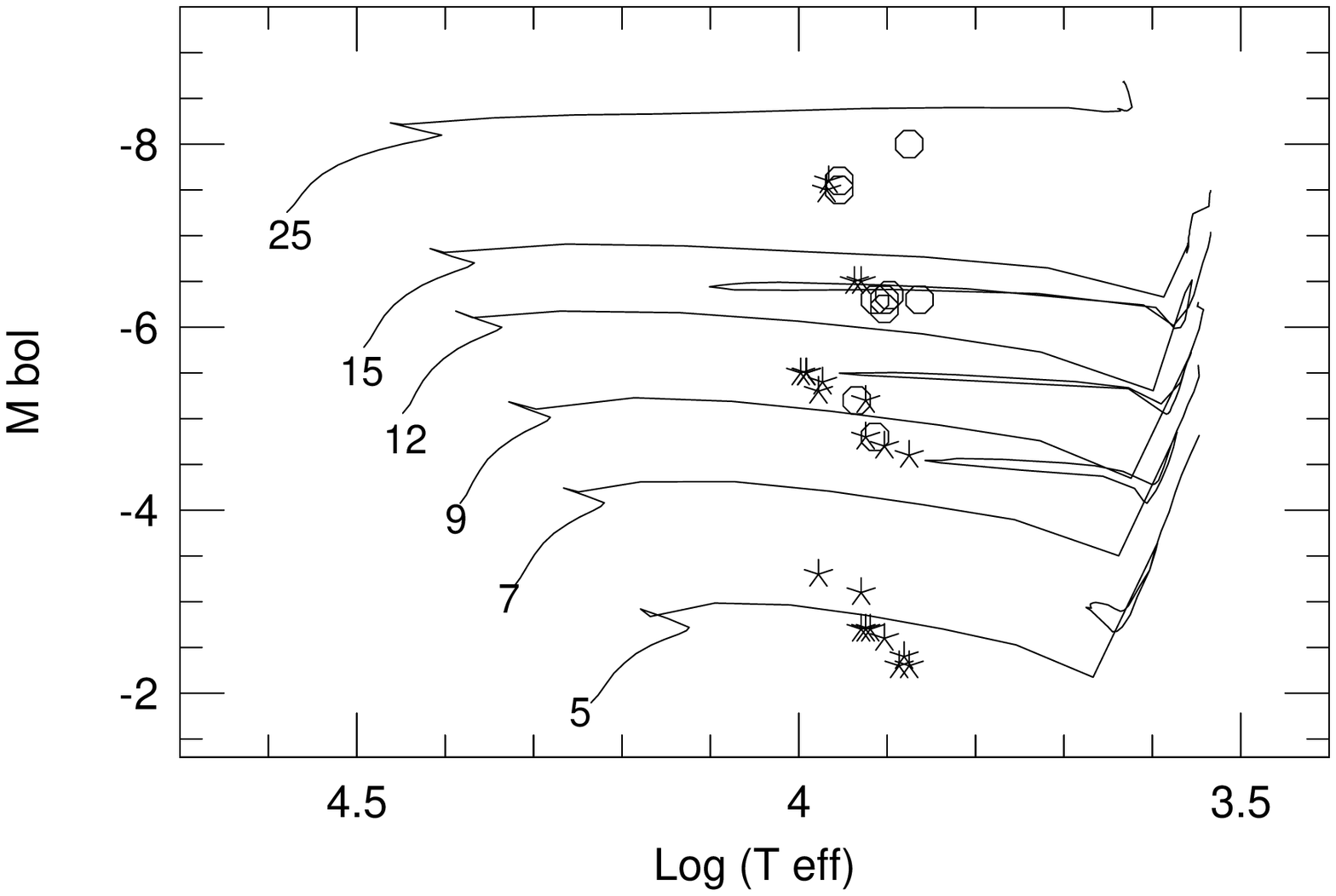}{3.0in}{0}{70}{70}{-270}{-270}
\caption{HR-diagram of SMC ({\it empty circles}) and 
Galactic ({\it stars}) A-supergiants analysed in
this paper and by Venn (1995a,b).   
Standard stellar evolution tracks 
from Schaller \etal (1992) are shown.  }
\end{figure}

\begin{figure}[t]
\plotfiddle{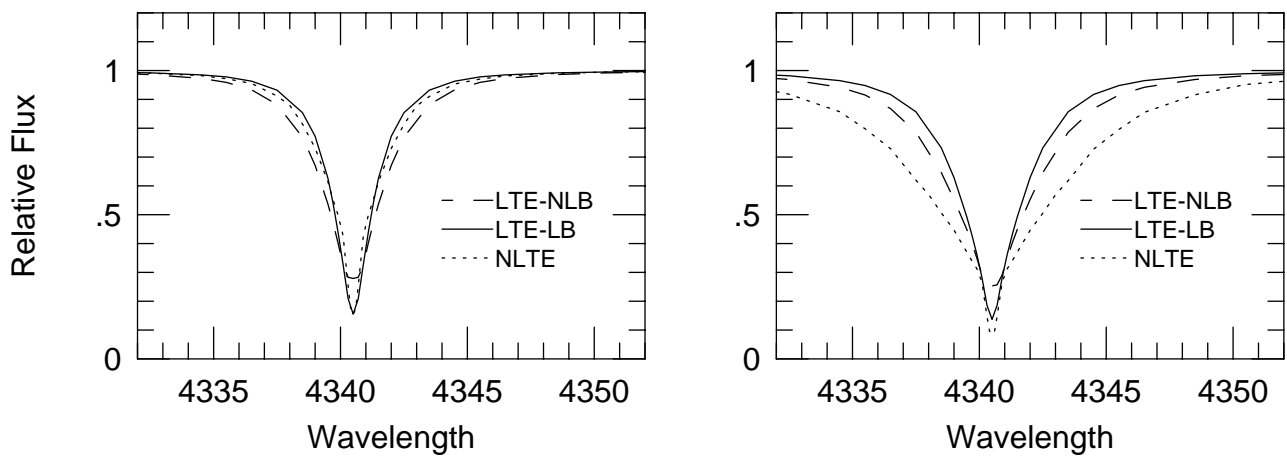}{3.0in}{0}{100}{100}{-300}{-550}
\caption{
Profiles of H$\gamma$ from three different sets of model atmospheres 
with \teff=8800~K and \logg=1.1 (left panel) and \logg=1.3 (right panel).    
Solid lines from Kurucz's ATLAS9 
including detailed ODFs.   Dashed lines from Kurucz's ATLAS8 with only 
H and He bound-free edges (no ODFs).  Dotted lines are from Kunze's NLTE 
models with H and He edges only (Kudritzki, private communications).
All models yield very similar H$\gamma$ line profiles at 
low gravity, but not at high gravity.}
\end{figure}

\begin{figure}[t]
\plotfiddle{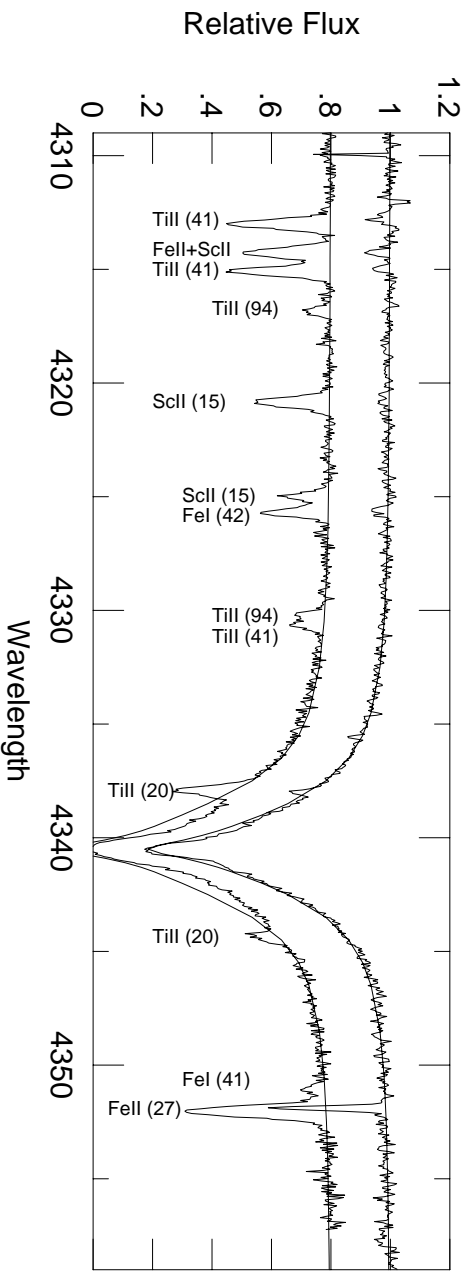}{3.0in}{90}{70}{70}{270}{-200}
\caption{Sample spectra of AV~298 (upper) and AV~442 (lower)
in the H$\gamma$ region.   Theoretical profiles from ATLAS9
using the atmospheric parameters listed in Table~1.}  
\end{figure}

\begin{figure}
\plotfiddle{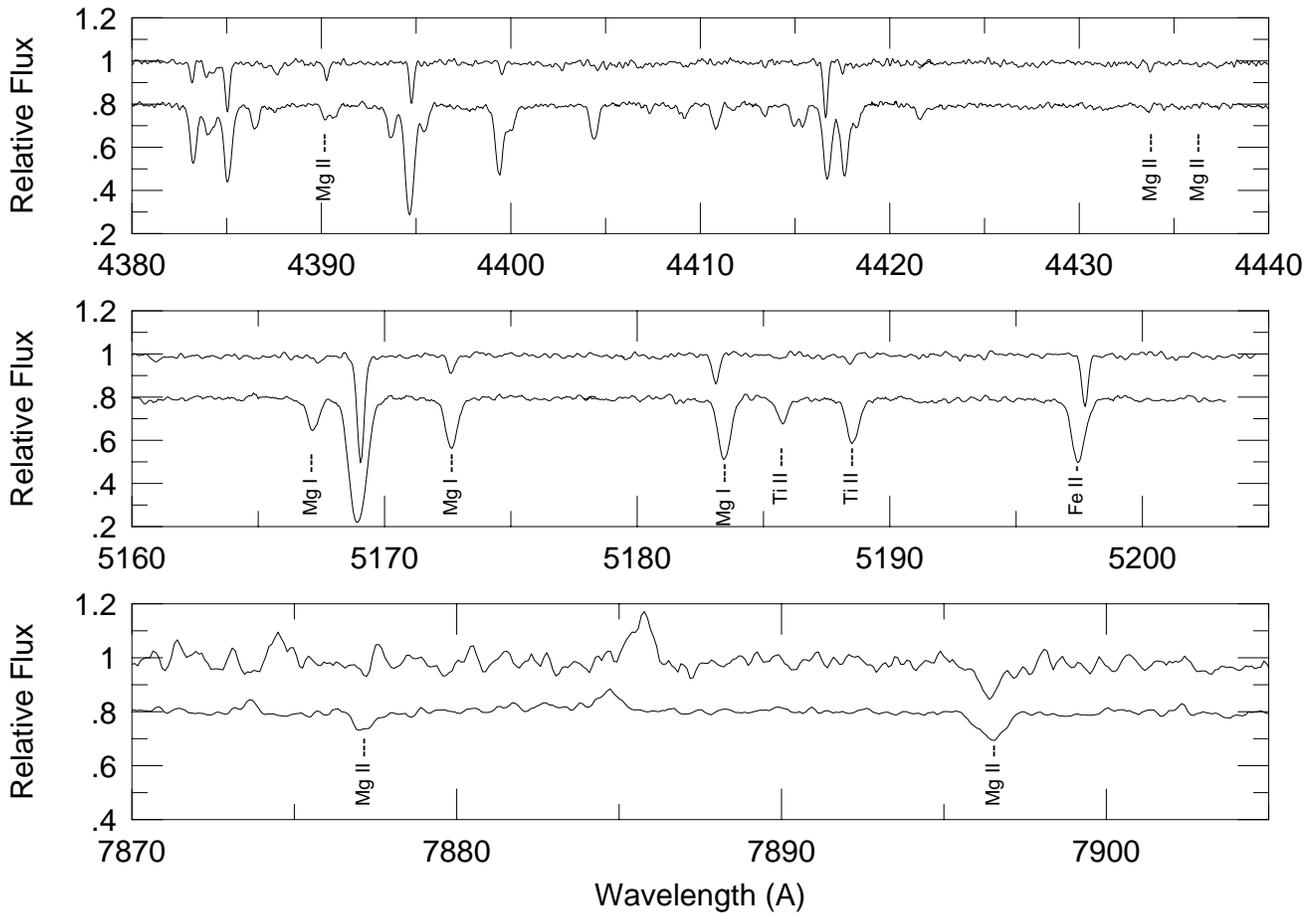}{5.0in}{90}{70}{70}{270}{-50}
\caption{Sample spectra of AV~298 (upper) and AV~442 (lower) 
to show some Mg~I and Mg~II lines used for atmospheric parameter 
and abundance determinations.  Spectra at 7900~\AA\ have been
divided by that of a hotter, rapidly-rotating star to remove 
telluric features.}
\end{figure}

\begin{figure}[t]
\plotfiddle{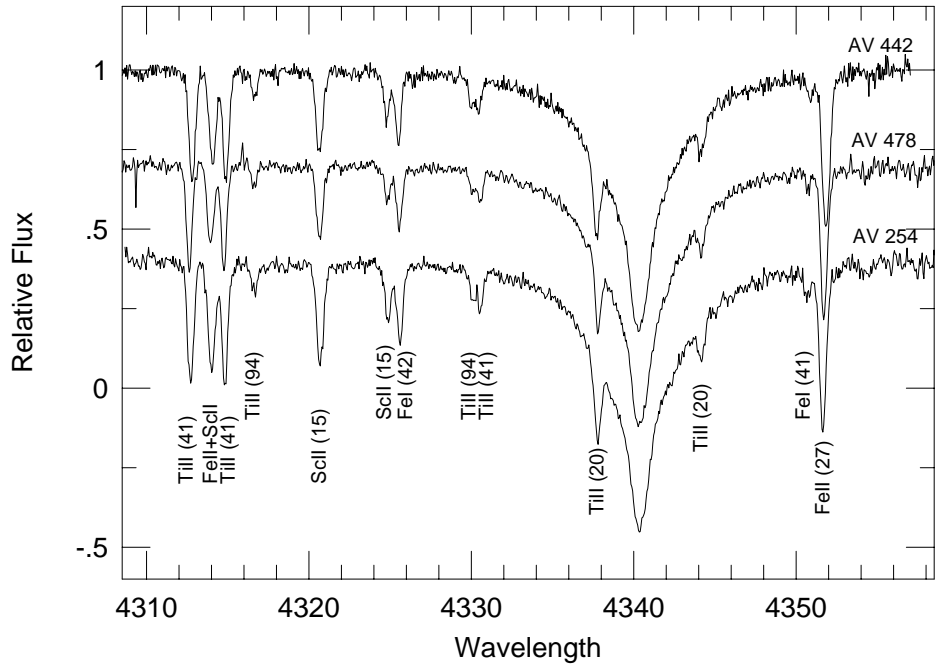}{3.4in}{90}{50}{50}{+200}{0}
\caption{ Spectra of three SMC A7\,Iab supergiants around H$\gamma$.   
Previously, AV~442 had been reported as a normal A3~Ia star, while the 
other two were reported as ``anomalous'', helium-rich A3~I stars 
partially based on a comparison of their hydrogen line profiles 
(Humphreys 1983, \etal 1991). }
\end{figure}

\begin{figure}
\plotfiddle{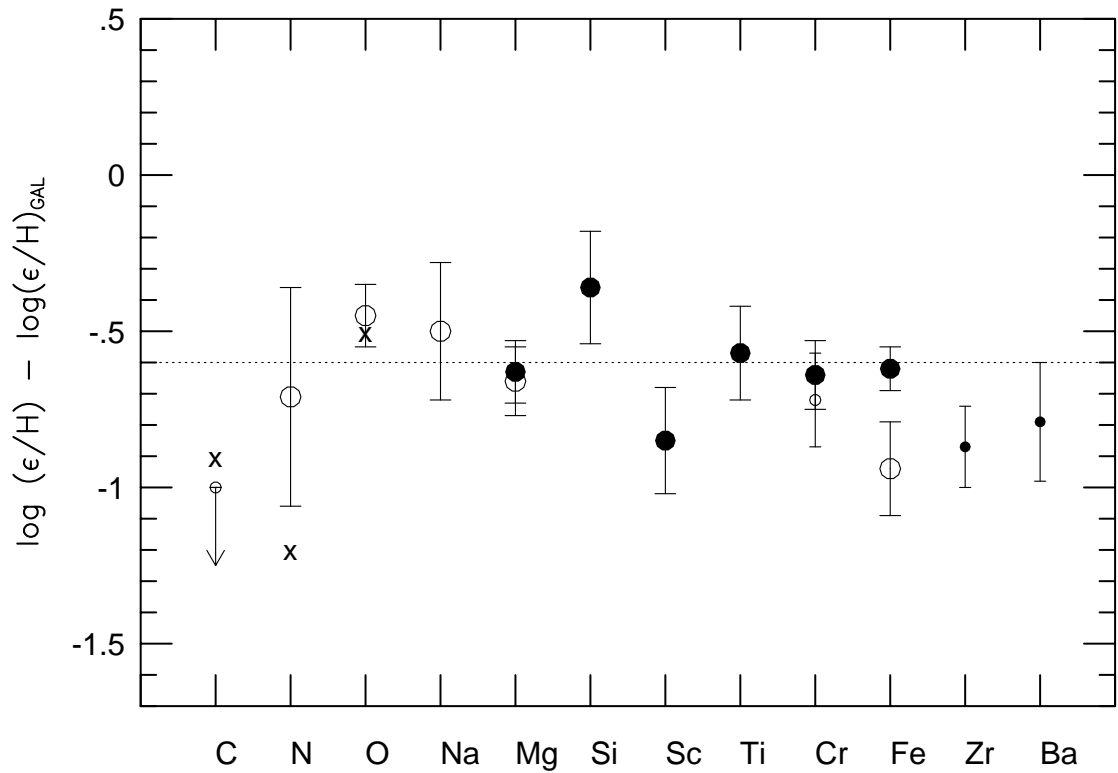}{4.0in}{90}{60}{60}{+200}{-30}
\caption{ The average abundances of SMC A-type supergiants 
relative to Galactic A-type supergiants. 
Exceptions include Na, Zr, and Ba, which are relative to
solar abundances (from Anders \& Grevesse 1989)  
since these elements are not well determined in the Galactic stars. 
Abundances from ions ({\it filled circles}) 
and neutrals ({\it hollow circles}) are noted separately; 
also, the size of the data points are related to the number 
of stars used in each average.   
The mean depletion of the metals ($\sim$0.6~dex) is noted 
by the dotted line.   
{\sl X}'s mark the SMC nebular abundances for CNO.}
\end{figure}

\begin{figure}[t]
\plotfiddle{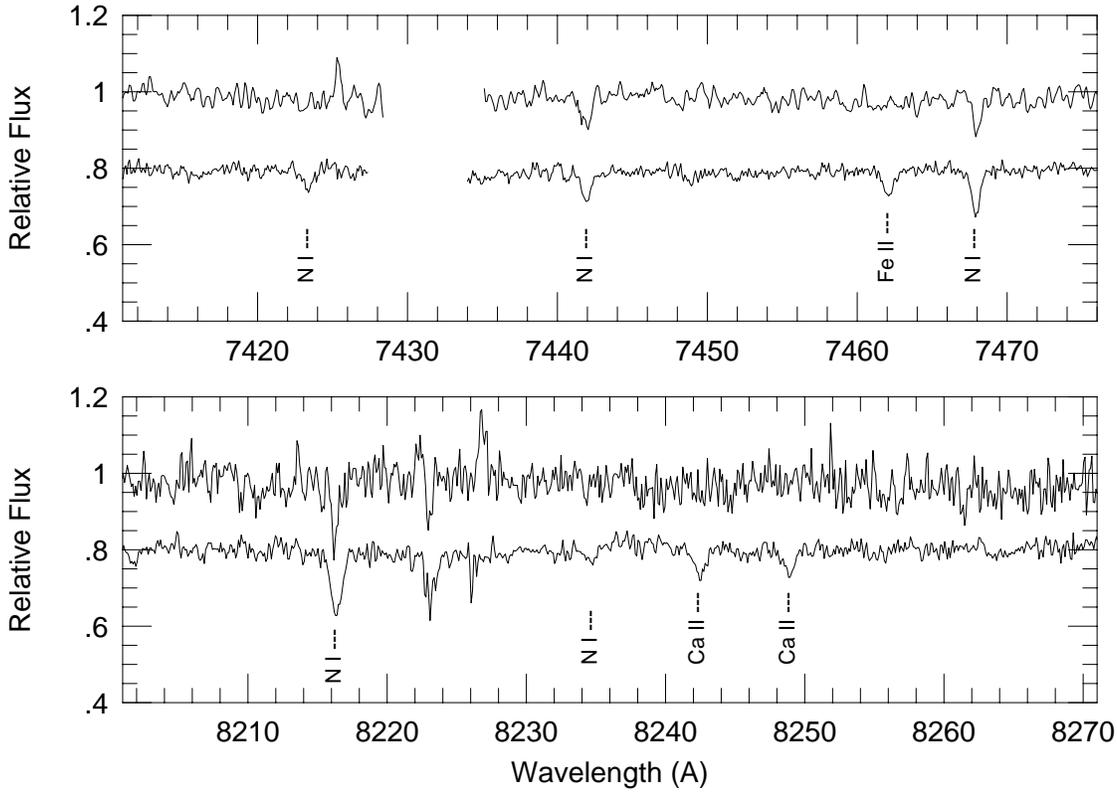}{4.0in}{90}{60}{60}{200}{-30}
\caption{Sample spectra of AV~298 (upper) and AV~442 (lower) 
to show N~I lines.   Spectra at 8200~\AA\ have been divided
by that of a hotter, rapidly-rotating star to remove 
telluric features.} 
\end{figure}

\begin{figure}[t]
\plotfiddle{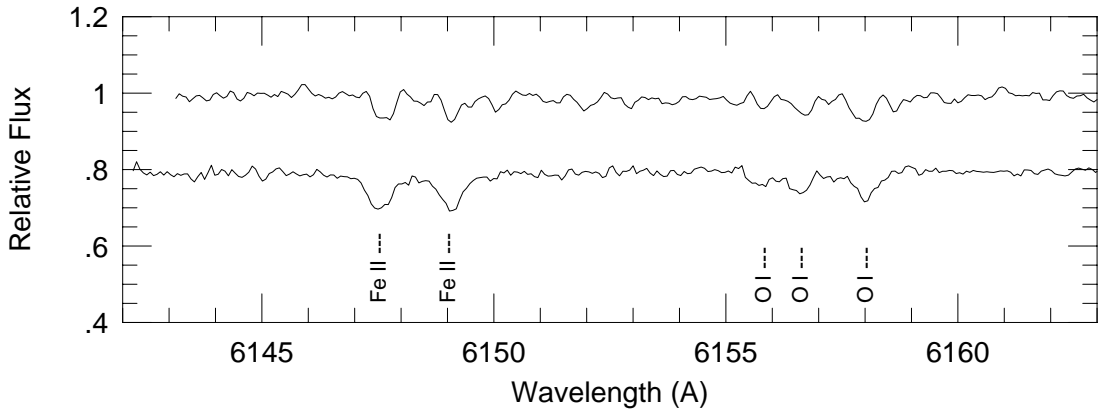}{3.0in}{90}{60}{60}{200}{-170}
\caption{Sample spectra of AV~298 (upper) and AV~442 (lower) 
to show O~I lines.}
\end{figure}

\begin{figure}
\plotfiddle{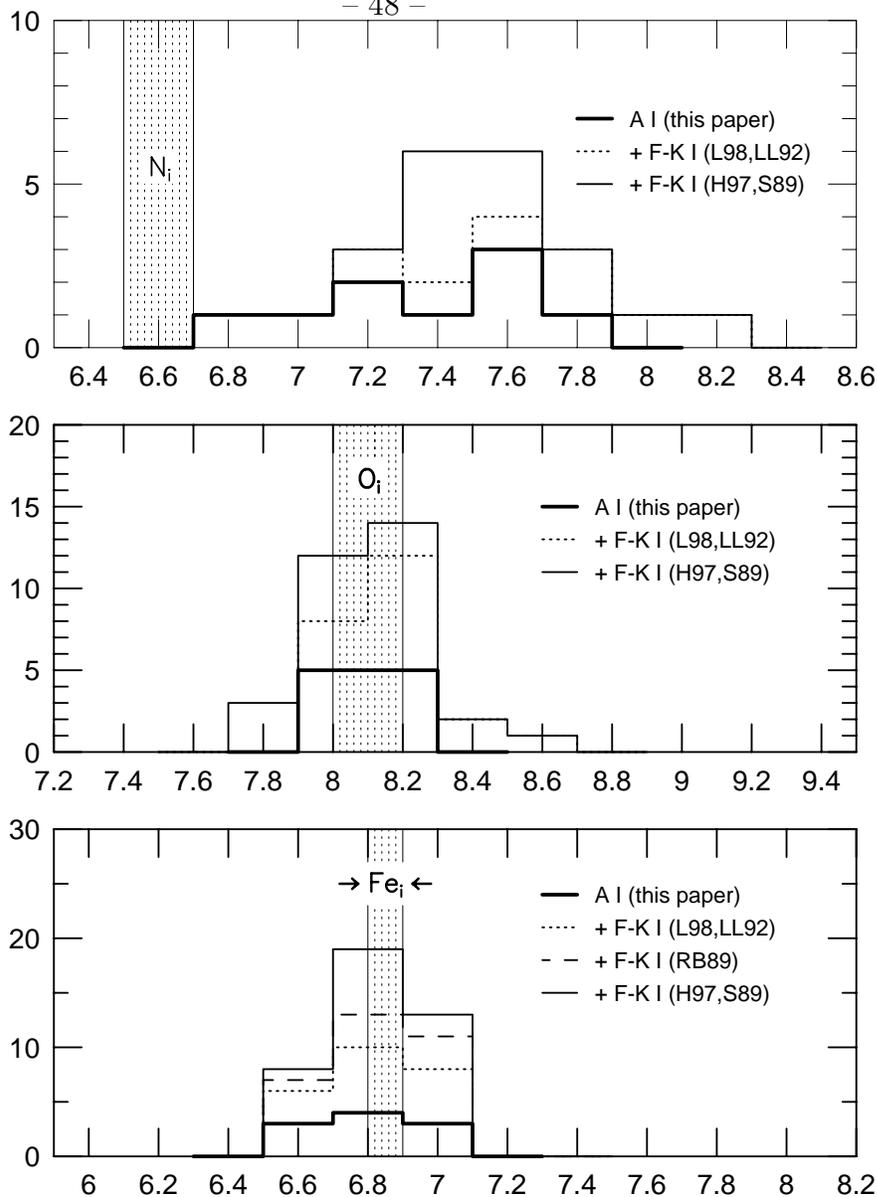}{5.0in}{0}{70}{70}{-200}{-50}
\caption{ Histograms of the N, O, and Fe abundances 
in A- to K-type supergiants in the SMC.  
The A-supergiant abundances (this paper) are graphed by a 
thick, solid line, and the F-K supergiant abundances 
(see text for references) are added on top for clarity. 
Initial, present-day, abundances are noted as grey regions.   
Notice that the range in N is much larger than O or Fe.}
\end{figure}

\begin{figure}
\plotfiddle{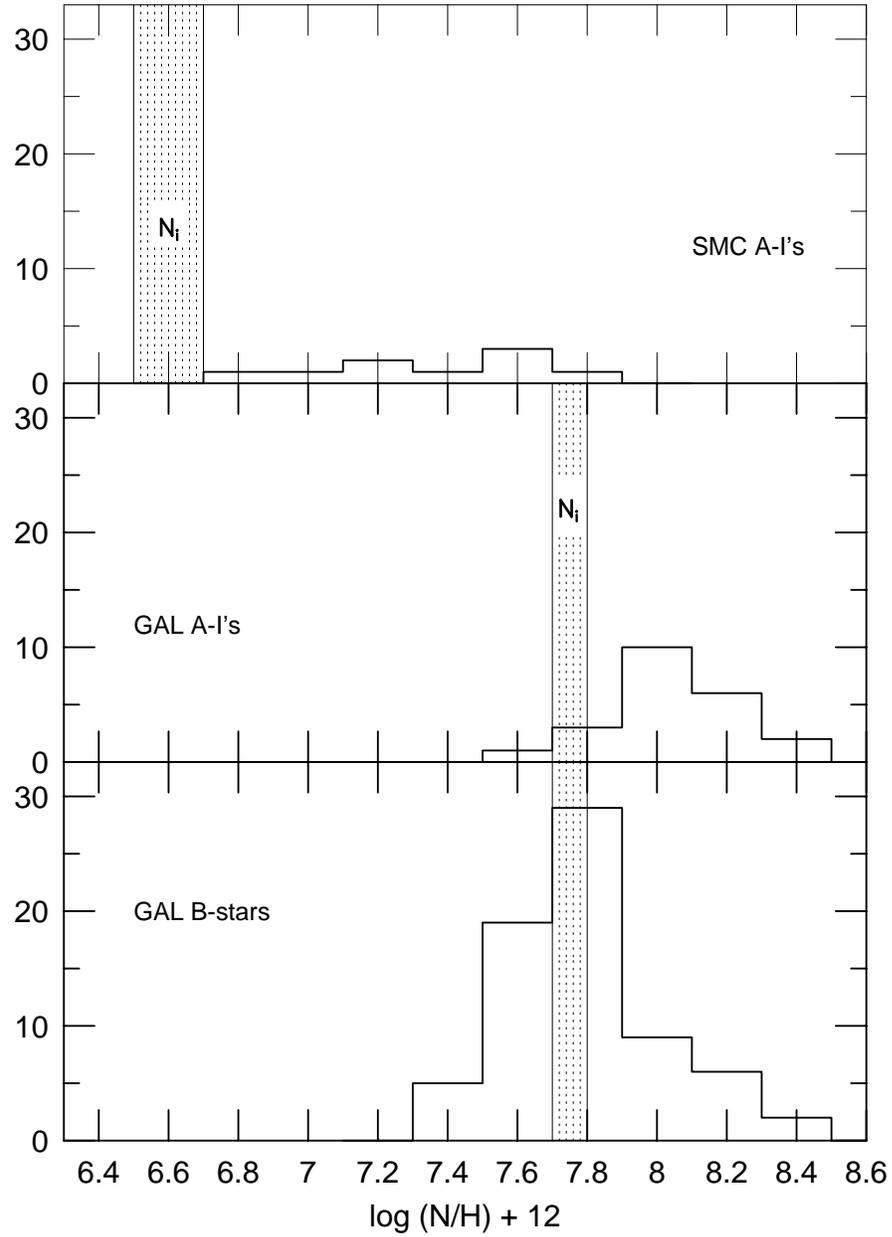}{5.0in}{0}{70}{70}{-200}{-50}
\caption{ Distribution of nitrogen abundances in the SMC and Galactic 
A-supergiants, and Galactic B-stars (see text for references). 
The initial abundances are marked.
Notice that the SMC stars are more enriched in N overall,
and yet some stars show very little enrichment. }
\end{figure}

\begin{figure}[t]
\plotfiddle{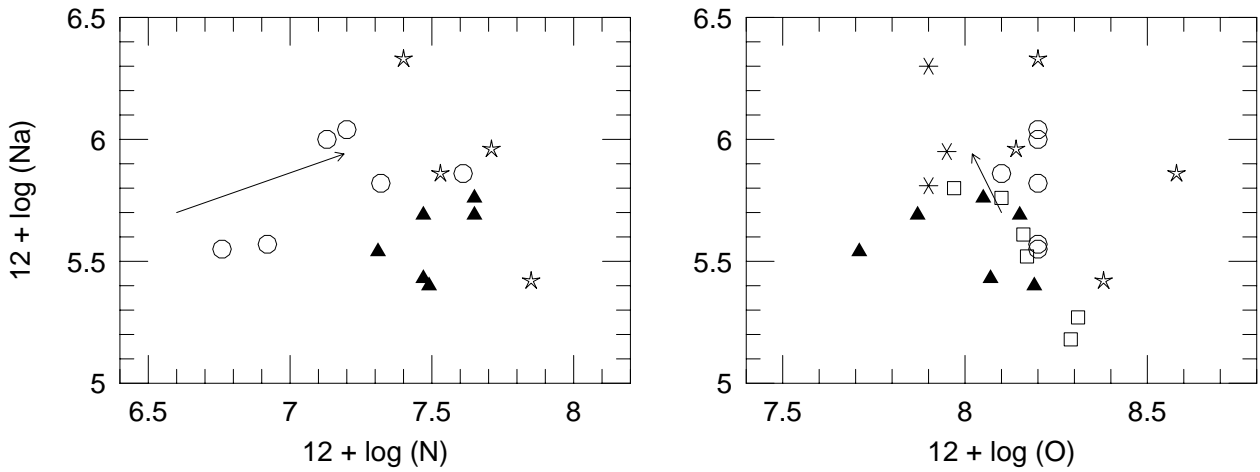}{3.5in}{90}{70}{70}{270}{-200}
\caption{Comparison of Na with N and O determinations in SMC
supergiants.   No statistically significant trends are found, suggesting 
that Na enrichments are not related to mixing of Ne-Na cycled gas, 
which would be accompanied by CNO-cycled gas.   Data is from this 
paper ({\it open circles}), Luck \etal (1998, {\it open squares}),
Luck \& Lambert (1992, {\it stars}), Hill (1997a,b, {\it filled
triangles}), and Spite \etal (1989a,b, {\it asterisks}).  Arrows indicate
predicted first dredge-up abundances for classical non-rotating stellar
evolution models (Heger 1998). }
\end{figure}

\end{document}